
\font\twelverm=cmr10  scaled 1200   \font\twelvei=cmmi10  scaled 1200
\font\twelvesy=cmsy10 scaled 1200   \font\twelveex=cmex10 scaled 1200
\font\twelvebf=cmbx10 scaled 1200   \font\twelvesl=cmsl10 scaled 1200
\font\twelvett=cmtt10 scaled 1200   \font\twelveit=cmti10 scaled 1200
\font\twelvesc=cmcsc10 scaled 1200  \font\twelvesf=cmsl10 scaled 1200
\skewchar\twelvei='177   \skewchar\twelvesy='60


\def\twelvepoint{\normalbaselineskip=12.4pt plus 0.1pt minus 0.1pt
  \abovedisplayskip 12.4pt plus 3pt minus 9pt
  \belowdisplayskip 12.4pt plus 3pt minus 9pt
  \abovedisplayshortskip 0pt plus 3pt
  \belowdisplayshortskip 7.2pt plus 3pt minus 4pt
  \smallskipamount=3.6pt plus1.2pt minus1.2pt
  \medskipamount=7.2pt plus2.4pt minus2.4pt
  \bigskipamount=14.4pt plus4.8pt minus4.8pt
  \def\rm{\fam0\twelverm}          \def\it{\fam\itfam\twelveit}%
  \def\sl{\fam\slfam\twelvesl}     \def\bf{\fam\bffam\twelvebf}%
  \def\mit{\fam 1}                 \def\cal{\fam 2}%
  \def\sc{\twelvesc}		   \def\tt{\twelvett}
  \def\sf{\twelvesf}
  \textfont0=\twelverm   \scriptfont0=\tenrm   \scriptscriptfont0=\sevenrm
  \textfont1=\twelvei    \scriptfont1=\teni    \scriptscriptfont1=\seveni
  \textfont2=\twelvesy   \scriptfont2=\tensy   \scriptscriptfont2=\sevensy
  \textfont3=\twelveex   \scriptfont3=\twelveex  \scriptscriptfont3=\twelveex
  \textfont\itfam=\twelveit
  \textfont\slfam=\twelvesl
  \textfont\bffam=\twelvebf \scriptfont\bffam=\tenbf
  \scriptscriptfont\bffam=\sevenbf
  \normalbaselines\rm}



\def\beginlinemode{\endmode
  \begingroup\parskip=0pt \obeylines\def\\{\par}\def\endmode{\par\endgroup}}
\def\beginparmode{\endmode
  \begingroup \def\endmode{\par\endgroup}}
\let\endmode=\par
{\obeylines\gdef\
{}}
\def\singlespace{\baselineskip=\normalbaselineskip}

\def\oneandahalfspace{\baselineskip=\normalbaselineskip
  \multiply\baselineskip by 3 \divide\baselineskip by 2}
\def\doublespace{\baselineskip=\normalbaselineskip \multiply\baselineskip by 2}

\newcount\firstpageno
\firstpageno=2
\footline={\ifnum\pageno<\firstpageno{\hfil}
\else{\hfil\twelverm\folio\hfil}\fi}
\def\toppageno{\global\footline={\hfil}\global\headline
  ={\ifnum\pageno<\firstpageno{\hfil}\else{\hfil\twelverm\folio\hfil}\fi}}
\let\rawfootnote=\footnote		
\def\footnote#1#2{{\rm\singlespace\parindent=0pt\parskip=0pt
  \rawfootnote{#1}{#2\hfill\vrule height 0pt depth 6pt width 0pt}}}
\def\raggedcenter{\leftskip=4em plus 12em \rightskip=\leftskip
  \parindent=0pt \parfillskip=0pt \spaceskip=.3333em \xspaceskip=.5em
  \pretolerance=9999 \tolerance=9999
  \hyphenpenalty=9999 \exhyphenpenalty=9999 }
\def\dateline{\rightline{\ifcase\month\or
  January\or February\or March\or April\or May\or June\or
  July\or August\or September\or October\or November\or December\fi
  \space\number\year}}
\def\received{\vskip 3pt plus 0.2fill
 \centerline{\sl (Received\space\ifcase\month\or
  January\or February\or March\or April\or May\or June\or
  July\or August\or September\or October\or November\or December\fi
  \qquad, \number\year)}}


\hsize=6.5truein
\hoffset=0truein
\vsize=8.9truein
\voffset=0truein
\parskip=\medskipamount
\def\\{\cr}
\twelvepoint		
\doublespace		
\overfullrule=0pt	


\newcount\timehour
\newcount\timeminute
\newcount\timehourminute
\def\daytime{\timehour=\time\divide\timehour by 60
  \timehourminute=\timehour\multiply\timehourminute by-60
  \timeminute=\time\advance\timeminute by \timehourminute
  \number\timehour:\ifnum\timeminute<10{0}\fi\number\timeminute}
\def\today{\number\day\space\ifcase\month\or Jan\or Feb\or Mar
  \or Apr\or May\or Jun\or Jul\or Aug\or Sep\or Oct\or
  Nov\or Dec\fi\space\number\year}




\def\title			
  {\null\vskip 3pt plus 0.2fill
   \beginlinemode \doublespace \raggedcenter \bf}

\def\author			
  {\vskip 3pt plus 0.2fill \beginlinemode
   \doublespace \raggedcenter}

\def\affil			
  {\vskip 3pt plus 0.1fill \beginlinemode
   \oneandahalfspace \raggedcenter \it}

\def\abstract			
  {\vskip 3pt plus 0.3fill \beginparmode \narrower
   \oneandahalfspace {\it  Abstract}:\  }

\def\endtopmatter		
  {\endpage			
   \body}

\def\body			
  {\beginparmode}		

\def\head#1{			
  \goodbreak\vskip 0.4truein	
  {\immediate\write16{#1}
   \raggedcenter {\sc #1} \par }
   \nobreak\vskip 0truein\nobreak}

\def\subhead#1{			
  \vskip 0.25truein		
  {\raggedcenter {\it #1} \par}
   \nobreak\vskip 0truein\nobreak}

\def\beneathrel#1\under#2{\mathrel{\mathop{#2}\limits_{#1}}}

\def\refto#1{$^{#1}$}		

\def\references			
  {\head{References}		
   \beginparmode
   \frenchspacing \parindent=0pt    
   \parskip=0pt \everypar{\hangindent=20pt\hangafter=1}}

\gdef\refis#1{\item{#1.\ }}			

\gdef\journal#1,#2,#3,#4.{		
    {\sl #1~}{\bf #2}, #3 (#4).}		

\def\prd{\journal Phys. Rev. D }

\def\pl{\journal Phys. Lett. }

\def\endreferences{\body}

\def\figurecaptions		
  {\endpage
   \beginparmode
   \head{Figure Captions}
}

\def\endpage			
  {\vfill\eject}

\def\endpaper			
  {\endmode\vfill\supereject}

\def\endit
  {\endpaper\end}


\def\ref#1{Ref.~#1}			
\def\Ref#1{Ref.~#1}			
\def\[#1]{[\cite{#1}]}
\def\cite#1{{#1}}
\def\(#1){(\call{#1})}
\def\call#1{{#1}}
\def\taghead#1{}
\def\frac#1#2{{#1 \over #2}}

\def\12{{1\over2}}

\def\ie{{\it i.e.,\ }}

\def\etc{{\it etc.\ }}

\def\sla{\raise.15ex\hbox{$/$}\kern-.57em}
\def\leaderfill{\leaders\hbox to 1em{\hss.\hss}\hfill}
\def\twiddle{\lower.9ex\rlap{$\kern-.1em\scriptstyle\sim$}}
\def\bigtwiddle{\lower1.ex\rlap{$\sim$}}
\def\gtwid{\mathrel{\raise.3ex\hbox{$>$\kern-.75em\lower1ex\hbox{$\sim$}}}}
\def\ltwid{\mathrel{\raise.3ex\hbox{$<$\kern-.75em\lower1ex\hbox{$\sim$}}}}
\def\square{\kern1pt\vbox{\hrule height 1.2pt\hbox{\vrule width 1.2pt\hskip 3pt
   \vbox{\vskip 6pt}\hskip 3pt\vrule width 0.6pt}\hrule height 0.6pt}\kern1pt}
\def\tdot#1{\mathord{\mathop{#1}\limits^{\kern2pt\ldots}}}

\def\pmb#1{\setbox0=\hbox{#1}%
  \kern-.025em\copy0\kern-\wd0
  \kern  .05em\copy0\kern-\wd0
  \kern-.025em\raise.0433em\box0 }


\def\gev{{\,\rm GeV}}

\catcode`@=11
\newcount\tagnumber\tagnumber=0

\immediate\newwrite\eqnfile
\newif\if@qnfile\@qnfilefalse
\def\write@qn#1{}
\def\writenew@qn#1{}
\def\w@rnwrite#1{\write@qn{#1}\message{#1}}
\def\@rrwrite#1{\write@qn{#1}\errmessage{#1}}

\def\taghead#1{\gdef\t@ghead{#1}\global\tagnumber=0}
\def\t@ghead{}

\expandafter\def\csname @qnnum-3\endcsname
  {{\t@ghead\advance\tagnumber by -3\relax\number\tagnumber}}
\expandafter\def\csname @qnnum-2\endcsname
  {{\t@ghead\advance\tagnumber by -2\relax\number\tagnumber}}
\expandafter\def\csname @qnnum-1\endcsname
  {{\t@ghead\advance\tagnumber by -1\relax\number\tagnumber}}
\expandafter\def\csname @qnnum0\endcsname
  {\t@ghead\number\tagnumber}
\expandafter\def\csname @qnnum+1\endcsname
  {{\t@ghead\advance\tagnumber by 1\relax\number\tagnumber}}
\expandafter\def\csname @qnnum+2\endcsname
  {{\t@ghead\advance\tagnumber by 2\relax\number\tagnumber}}
\expandafter\def\csname @qnnum+3\endcsname
  {{\t@ghead\advance\tagnumber by 3\relax\number\tagnumber}}

\def\equationfile{%
  \@qnfiletrue\immediate\openout\eqnfile=\jobname.eqn%
  \def\write@qn##1{\if@qnfile\immediate\write\eqnfile{##1}\fi}
  \def\writenew@qn##1{\if@qnfile\immediate\write\eqnfile
    {\noexpand\tag{##1} = (\t@ghead\number\tagnumber)}\fi}
}

\def\callall#1{\xdef#1##1{#1{\noexpand\call{##1}}}}
\def\call#1{\each@rg\callr@nge{#1}}

\def\each@rg#1#2{{\let\thecsname=#1\expandafter\first@rg#2,\end,}}
\def\first@rg#1,{\thecsname{#1}\apply@rg}
\def\apply@rg#1,{\ifx\end#1\let\next=\relax%
\else,\thecsname{#1}\let\next=\apply@rg\fi\next}

\def\callr@nge#1{\calldor@nge#1-\end-}
\def\callr@ngeat#1\end-{#1}
\def\calldor@nge#1-#2-{\ifx\end#2\@qneatspace#1 %
  \else\calll@@p{#1}{#2}\callr@ngeat\fi}
\def\calll@@p#1#2{\ifnum#1>#2{\@rrwrite{Equation range #1-#2\space is bad.}
\errhelp{If you call a series of equations by the notation M-N, then M and
N must be integers, and N must be greater than or equal to M.}}\else%
{\count0=#1\count1=#2\advance\count1 by1\relax\expandafter\@qncall\the\count0,%
  \loop\advance\count0 by1\relax%
    \ifnum\count0<\count1,\expandafter\@qncall\the\count0,%
  \repeat}\fi}

\def\@qneatspace#1#2 {\@qncall#1#2,}
\def\@qncall#1,{\ifunc@lled{#1}{\def\next{#1}\ifx\next\empty\else
  \w@rnwrite{Equation number \noexpand\(>>#1<<) has not been defined yet.}
  >>#1<<\fi}\else\csname @qnnum#1\endcsname\fi}

\let\eqnono=\eqno
\def\eqno(#1){\tag#1}
\def\tag#1$${\eqnono(\displayt@g#1 )$$}

\def\aligntag#1\endaligntag
  $${\gdef\tag##1\\{&(##1 )\cr}\eqalignno{#1\\}$$
  \gdef\tag##1$${\eqnono(\displayt@g##1 )$$}}

\def\eqalignno#1{\displ@y \tabskip\centering
  \halign to\displaywidth{\hfil$\displaystyle{##}$\tabskip\z@skip
    &$\displaystyle{{}##}$\hfil\tabskip\centering
    &\llap{$\displayt@gpar##$}\tabskip\z@skip\crcr
    #1\crcr}}

\def\displayt@gpar(#1){(\displayt@g#1 )}

\def\displayt@g#1 {\rm\ifunc@lled{#1}\global\advance\tagnumber by1
        {\def\next{#1}\ifx\next\empty\else\expandafter
        \xdef\csname @qnnum#1\endcsname{\t@ghead\number\tagnumber}\fi}%
  \writenew@qn{#1}\t@ghead\number\tagnumber\else
        {\edef\next{\t@ghead\number\tagnumber}%
        \expandafter\ifx\csname @qnnum#1\endcsname\next\else
        \w@rnwrite{Equation \noexpand\tag{#1} is a duplicate number.}\fi}%
  \csname @qnnum#1\endcsname\fi}

\def\ifunc@lled#1{\expandafter\ifx\csname @qnnum#1\endcsname\relax}

\let\@qnend=\end\gdef\end{\if@qnfile
\immediate\write16{Equation numbers written on []\jobname.EQN.}\fi\@qnend}

\catcode`@=12

\catcode`@=11
\newcount\r@fcount \r@fcount=0
\newcount\r@fcurr
\immediate\newwrite\reffile
\newif\ifr@ffile\r@ffilefalse
\def\w@rnwrite#1{\ifr@ffile\immediate\write\reffile{#1}\fi\message{#1}}

\def\writer@f#1>>{}
\def\referencefile{
  \r@ffiletrue\immediate\openout\reffile=\jobname.ref%
  \def\writer@f##1>>{\ifr@ffile\immediate\write\reffile%
    {\noexpand\refis{##1} = \csname r@fnum##1\endcsname = %
     \expandafter\expandafter\expandafter\strip@t\expandafter%
     \meaning\csname r@ftext\csname r@fnum##1\endcsname\endcsname}\fi}%
  \def\strip@t##1>>{}}

\def\citeall#1{\xdef#1##1{#1{\noexpand\cite{##1}}}}
\def\cite#1{\each@rg\citer@nge{#1}}

\def\each@rg#1#2{{\let\thecsname=#1\expandafter\first@rg#2,\end,}}
\def\first@rg#1,{\thecsname{#1}\apply@rg}	
\def\apply@rg#1,{\ifx\end#1\let\next=\relax
\else,\thecsname{#1}\let\next=\apply@rg\fi\next}

\def\citer@nge#1{\citedor@nge#1-\end-}	
\def\citer@ngeat#1\end-{#1}
\def\citedor@nge#1-#2-{\ifx\end#2\r@featspace#1 
  \else\citel@@p{#1}{#2}\citer@ngeat\fi}	
\def\citel@@p#1#2{\ifnum#1>#2{\errmessage{Reference range #1-#2\space is bad.}
    \errhelp{If you cite a series of references by the notation M-N, then M and
    N must be integers, and N must be greater than or equal to M.}}\else%
{\count0=#1\count1=#2\advance\count1 by1\relax\expandafter\r@fcite\the\count0,%
  \loop\advance\count0 by1\relax
    \ifnum\count0<\count1,\expandafter\r@fcite\the\count0,%
  \repeat}\fi}

\def\r@featspace#1#2 {\r@fcite#1#2,}	
\def\r@fcite#1,{\ifuncit@d{#1}		
    \expandafter\gdef\csname r@ftext\number\r@fcount\endcsname%
    {\message{Reference #1 to be supplied.}\writer@f#1>>#1 to be supplied.\par
     }\fi%
  \csname r@fnum#1\endcsname}

\def\ifuncit@d#1{\expandafter\ifx\csname r@fnum#1\endcsname\relax%
\global\advance\r@fcount by1%
\expandafter\xdef\csname r@fnum#1\endcsname{\number\r@fcount}}

\let\r@fis=\refis			
\def\refis#1#2#3\par{\ifuncit@d{#1}
    \w@rnwrite{Reference #1=\number\r@fcount\space is not cited up to now.}\fi%
  \expandafter\gdef\csname r@ftext\csname r@fnum#1\endcsname\endcsname%
  {\writer@f#1>>#2#3\par}}

\def\r@ferr{\endreferences\errmessage{I was expecting to see
\noexpand\endreferences before now;  I have inserted it here.}}
\let\r@ferences=\references
\def\references{\r@ferences\def\endmode{\r@ferr\par\endgroup}}

\let\endr@ferences=\endreferences
\def\endreferences{\r@fcurr=0
  {\loop\ifnum\r@fcurr<\r@fcount
    \advance\r@fcurr by 1\relax\expandafter\r@fis\expandafter{\number\r@fcurr}%
    \csname r@ftext\number\r@fcurr\endcsname%
  \repeat}\gdef\r@ferr{}\endr@ferences}


\let\r@fend=\endpaper\gdef\endpaper{\ifr@ffile
\immediate\write16{Cross References written on []\jobname.REF.}\fi\r@fend}

\catcode`@=12

\citeall\refto		
\citeall\ref		%
\citeall\Ref		%

\catcode`@=11
\newwrite\tocfile\openout\tocfile=\jobname.toc
\newlinechar=`^^J
\write\tocfile{\string\input\space jnl^^J
  \string\pageno=-1\string\firstpageno=-1000\string\singlespace
  \string\null\string\vfill\string\centerline{TABLE OF CONTENTS}^^J
  \string\vskip 0.5 truein\string\rightline{\string\underbar{Page}}\smallskip}

\def\tocitem#1{
  \t@cskip{#1}\bigskip}
\def\tocitemitem#1{
  \t@cskip{\quad#1}\medskip}
\def\tocitemitemitem#1{
  \t@cskip{\qquad#1}\smallskip}
\def\tocitemall#1{
  \xdef#1##1{#1{##1}\noexpand\tocitem{##1}}}
\def\tocitemitemall#1{
  \xdef#1##1{#1{##1}\noexpand\tocitemitem{##1}}}
\def\tocitemitemitemall#1{
  \xdef#1##1{#1{##1}\noexpand\tocitemitemitem{##1}}}

\def\t@cskip#1#2{
  \write\tocfile{\string#2\string\line{^^J
  #1\string\leaderfill\space\number\folio}}}

%

\def\t@cproduce{
  \write\tocfile{\string\vfill\string\vfill\string\supereject\string\end}
  \closeout\tocfile
  \immediate\write16{Table of Contents written on []\jobname.TOC.}}
\def\mbox#1#2{\vcenter{\hrule width#1in\hbox{\vrule height#2in
     \hskip#1in\vrule height#2in}\hrule width#1in}}

\let\t@cend=\endpaper\def\endpaper{\t@cproduce\t@cend}

\catcode`@=12

\tocitemall\head		

%
%
%

\newbox\tabbox

\newif\ifthead
\global\theadfalse


%

\newcount\tablenum
\global\tablenum=0


\def\table#1#2#3{\topinsert
\vskip3pc
\setbox\tabbox=\vtop\bgroup
\global\advance\tablenum by 1
\hsize #3
\vbox{\noindent{\bf Table \the\tablenum.}\quad #1}
\vskip 6pt
\hrule width\hsize
\vskip 6pt
\parskip0pt
\halign{%
##\hfill%
\hskip4pt
&&
##\hfill\hskip4pt\cr%
#2\cr}
\hrule width\hsize
\egroup
\line{\hskip\leftskip\hfill\copy\tabbox\hfill}
\endinsert}

\def\midtable#1#2#3{\vbox{
\vskip1pc\setbox\tabbox=\vtop\bgroup
\global\advance\tablenum by 1
\hsize #3
\vbox{\noindent{\bf Table \the\tablenum.}\quad #1}
\vskip 6pt
\hrule width\hsize
\vskip 6pt
\parskip0pt
\halign{##\hfill\hskip4pt&&##\hfill\hskip4pt\cr
#2\cr}
\hrule width\hsize
\egroup
\line{\hskip\leftskip\hfill\copy\tabbox\hfill}
\vskip1pc}}
%


\font\figurefont cmr9
\font\PalatinoNineIt cmti9
\font\foliofont cmbx9  
\font\ninei=cmmi9
\font\ninesy=cmsy9

\def\ninepoint{\def\rm{\fam0\figurefont}
\textfont0=\figurefont \scriptfont0=\sevenrm\scriptscriptfont0=\fiverm
\textfont1=\ninei  \scriptfont1=\seveni \scriptscriptfont1=\fivei
\textfont2=\ninesy \scriptfont2=\sevensy \scriptscriptfont2=\fivesy
\textfont3=\tenex \scriptfont3=\tenex \scriptscriptfont3=\tenex
\textfont\bffam=\foliofont\def\bf{\fam\bffam\foliofont}
\textfont\slfam=\PalatinoNineIt\def\sl{\fam\slfam\PalatinoNineIt}
\textfont\itfam=\PalatinoNineIt\def\it{\fam\itfam\PalatinoNineIt}
\baselineskip11pt\rm}

\newcount\fignum
\global\fignum=0

\newcount\figtblnum
\global\figtblnum=0

\def\figure#1#2{
\topinsert\bgroup\global\advance\fignum by 1
\ninepoint
\parskip0pt
\vglue42pt
\centerline{\vbox to #1{}}
\vskip6pt\baselineskip11pt
\noindent{\bf Figure \the\fignum\enskip}{\rm #2}\hfil%
\vskip18pt
\egroup
\vglue-\topskip
\endinsert
}

\def\midfigure#1#2{
\bgroup
\ninepoint
\global\advance\fignum by 1\parskip0pt
\vglue42pt
\centerline{\vbox to #1{}}
\vskip6pt\baselineskip11pt
\noindent{\bf Figure \the\fignum\enskip}{\rm #2}\hfil%
\vskip18pt
\egroup
}

\def\figtbl#1#2{
\topinsert\bgroup\global\advance\figtblnum by 1
\ninepoint
\parskip0pt
\vglue42pt
\centerline{\vbox to #1{}}
\vskip6pt\baselineskip11pt
\noindent{\bf Table \the\figtblnum\enskip}{\rm #2}\hfil%
\vskip18pt
\egroup
\vglue-\topskip
\endinsert
}


\def\leaderfill{{\leaders\hbox to 1em{\hss.\hss}\hfill}}
\def\refto#1{$^{#1}$}         
\gdef\journal#1,#2,#3,#4.{         
    {\sl #1~}{\bf #2}, #3 (#4).}        

\def\prd{\journal Phys. Rev. D }

\def\pl{\journal Phys. Lett., }

\def\12{{1\over2}}
\def\ifmath#1{\relax\ifmmode #1\else $#1$\fi}

\def\frac#1/#2{\leavevmode\kern.1em\raise.5ex\hbox{\the\scriptfont0
         #1}\kern-.1em/\kern-.15em\lower.25ex\hbox{\the\scriptfont0 #2}}


\def\ie{{\it i.e.,\ }}

\def\etc{{\it etc.\ }}

\def\Dsl{\raise.15ex\hbox{$/$}\kern-.57em\hbox{$D$}}
\def\Esl{\raise.15ex\hbox{$/$}\kern-.57em\hbox{$E$}}
\def\Desl{\raise.15ex\hbox{$/$}\kern-.57em\hbox{$\Delta$}}
\def\pasl{{\raise.15ex\hbox{$/$}\kern-.57em\hbox{$\partial$}}}
\def\nasl{\raise.15ex\hbox{$/$}\kern-.57em\hbox{$\nabla$}}
\def\rasl{\raise.15ex\hbox{$/$}\kern-.57em\hbox{$\rightarrow$}}
\def\lasl{\raise.15ex\hbox{$/$}\kern-.57em\hbox{$\lambda$}}
\def\asl{\raise.15ex\hbox{$/$}\kern-.57em\hbox{$a$}}
\def\bsl{\raise.15ex\hbox{$/$}\kern-.57em\hbox{$b$}}
\def\csl{\raise.15ex\hbox{$/$}\kern-.57em\hbox{$c$}}
\def\dsl{\raise.15ex\hbox{$/$}\kern-.57em\hbox{$d$}}
\def\hsl{\raise.15ex\hbox{$/$}\kern-.57em\hbox{$h$}}
\def\psl{\raise.15ex\hbox{$/$}\kern-.57em\hbox{$p$}}
\def\Psl{\raise.15ex\hbox{$/$}\kern-.57em\hbox{$P$}}
\def\Ksl{\raise.15ex\hbox{$/$}\kern-.57em\hbox{$K$}}
\def\ksl{\raise.15ex\hbox{$/$}\kern-.57em\hbox{$k$}}
\def\qsl{\raise.15ex\hbox{$/$}\kern-.57em\hbox{$q$}}
\def\gtwid{\raise.3ex\hbox{$>$\kern-.75em\lower1ex\hbox{$\sim$}}}
\def\ltwid{\raise.3ex\hbox{$<$\kern-.75em\lower1ex\hbox{$\sim$}}}

\def\boxit#1{\vbox{\hrule\hbox{\vrule\kern3pt
      \vbox{\kern3pt#1\kern3pt}\kern3pt\vrule}\hrule}}
\def\gev{{\rm\,GeV}}
\def\tev{{\rm\,TeV}}
\def\mev{{\rm\,MeV}}




\def\De{{\Delta}}


\def\qbar{{\bar q}}

\def\tbar{{\bar t}}

\def\nubar{{\bar\nu}}

\def\khat{{\hat k}}

\def\beneathrel#1\under#2{\mathrel{\mathop{#2}\limits_{#1}}}
\def\part{\partial}
\def\low#1{\lower.5ex\hbox{${}_#1$}}

\def\partt{\raise.15ex\hbox{$\widetilde$}{\kern-.37em\hbox{$\partial$}}}

\def\su3{{$SU(3)$}}
\def\Li2{{{\rm Li}_2}}

\def\q2{{Q^2}}
\def\Q2{{Q^2}}


\rightline{UFIFT-HEP-93-3}
\rightline{March 1993}

\edef\tfontsize{scaled\magstep3} 
\font\titlerm=cmr10 \tfontsize \font\titlerms=cmr7 \tfontsize
\font\titlermss=cmr5 \tfontsize \font\titlei=cmmi10 \tfontsize
\font\titleis=cmmi7 \tfontsize \font\titleiss=cmmi5 \tfontsize
\font\titlesy=cmsy10 \tfontsize \font\titlesys=cmsy7 \tfontsize
\font\titlesyss=cmsy5 \tfontsize \font\titleit=cmti10 \tfontsize
\skewchar\titlei='177 \skewchar\titleis='177 \skewchar\titleiss='177
\skewchar\titlesy='60 \skewchar\titlesys='60 \skewchar\titlesyss='60
\def\titlefont{\def\rm{\fam0\titlerm}
\textfont0=\titlerm \scriptfont0=\titlerms \scriptscriptfont0=\titlermss
\textfont1=\titlei \scriptfont1=\titleis \scriptscriptfont1=\titleiss
\textfont2=\titlesy \scriptfont2=\titlesys \scriptscriptfont2=\titlesyss
\textfont\itfam=\titleit \def\it{\fam\itfam\titleit}\rm}

\title{{\titlefont Enhancing the heavy Higgs signal
with jet-jet profile \
cuts\footnote{${}^\dagger$}{\twelverm Supported \
in part by U.S.~Department of Energy grant DE--FG05--86ER--40272.}}}

\author{R. D. Field${}^{*}$ and P. A. \
Griffin\footnote{${}^{*}$}{E-mail (Internet): \
rfield@ufhepa.phys.ufl.edu, pgriffin@ufhepa.phys.ufl.edu.}}
\affil
Institute for Fundamental Theory, Department of Physics
University of Florida, Gainesville, FL 32611

\abstract
The jet-jet profile, or detailed manner, in which transverse energy and mass
are distributed around the jet-jet system resulting from the hadronic decay
of a $Z$ boson in the process Higgs$\to ZZ$ at a proton-proton collider
energy of $40\tev$ is carefully examined.  Two observables are defined that
can be used to help distinguish the $\ell^+\ell^-$-jet-jet signal from Higgs
decay from the ``ordinary'' QCD background arising from the large
transverse momentum production of single $Z$ bosons plus the associated
jets.  By making cuts on these observables, signal to background
enhancement factors greater than $100$ can be obtained.

\endtopmatter

\head{I. Introduction}

The great challenge at hadron colliders is to disentangle any new physics
that may be present from the ``ordinary'' QCD background.  An important
final state consists of a large transverse momentum charged lepton pair plus
two accompanying jets (\ie $\ell^+\ell^-jj$).  It is one of the relevant
signals
for the production of a Higgs particle and its subsequent decay into $ZZ$
with one $Z$ decaying leptonically and the other $Z$ decaying
hadronically into a $q\qbar$ pair which then manifests itself as a pair of jets
[1,2]. Unfortunately, too often the large transverse momentum production
of single $Z$ bosons plus the associated jets mimic the Higgs signal.  Once
one requires the $Z$ boson to have a large transverse momentum by
demanding a large $P_T$ lepton pair, one has forced the background to
have a large $P_T$ ``away-side'' quark or gluon via subprocesses like
$qg\to Zq$ or $q\qbar\to Zg$.  This away-side parton often fragments via
gluon bremsstrahlung, producing away-side jet pairs which resemble the
signal. However, the signal jet pair and the background jet pair have quite
different origins.  The former arises from the decay of a color singlet $Z$
boson while the later is produced in a color non-singlet ``parton shower''.
We examine in detail the jet-jet profile, or precise manner, in which
transverse energy and mass are distributed around this jet-jet system.  Two
observables are defined that can be used to help distinguish the Higgs
signal from the QCD background.  By making cuts on these observables,
signal to background enhancement factors greater than $100$ can be
obtained, where the enhancement factor is defined to be the percentage of
signal divided by the percentage of background surviving a given set of
cuts.

We begin in Section II by discussing the event generation and selection
criteria.  The jet-jet profile analysis is presented in Section III with
Section
IV being reserved for summary and conclusions.
\vfill
\figtbl{3.5in}{(a) Single jet and (b) jet pair definitions.}

\head{II. Event Generation and Cuts}

ISAJET version 6.50 is used to generate $50,000$ Higgs bosons with a
mass of $800\gev$ in $40\tev$ proton-proton collisions.  The width of the
Higgs is $261\gev$ and it is generated over a mass range of $600$ to
$1000\gev$ and forced to decay into two $Z$ bosons.  For $800\gev$
Higgs at $40\tev$, the dominate subprocesses are $gg\to H$ with $35\%$
of the cross section and $t\tbar\to H$ at $36\%$ ($m_t=140\gev$). $WW$
and $ZZ$ fusion make up $21\%$ and $7\%$ of the signal, respectively.
In addition, $100,000$ $ZZ$ continuum events (\ie $q\qbar\to ZZ$) and
$400,000$ single $Z$ boson events are generated with the hard-scattering
transverse momentum of the $Z$, $\khat_T$, in the range
$150\le\khat_T\le 1000\gev$.  Single $Z$ bosons are produced at large
transverse momentum via the ``ordinary'' QCD subprocesses $qg\to Zq$,
$\qbar g\to Z\qbar$, and $q\qbar\to Zg$.  These subprocesses, of course,
generate addition gluons via bremsstrahlung off both incident and outgoing
color non-singlet partons, resulting in multiparton final states which
subsequently fragment into hadrons, and is referred to as the $Z+$jets
background.

Events are analyzed by dividing the solid angle into ``calorimeter'' cells
having size $\De\eta\De\phi=0.1\times7.5^\circ$, where $\eta$ and $\phi$
are the pseudorapidity and azimuthal angle, respectively.  A single cell has
an energy (the sum of the energies of all the particles that hit the cell {\it
excluding} neutrinos) and a direction given by the coordinates of the center
of the cell.  From this the transverse energy of each cell is computed from
the cell energy and direction.  Large transverse momentum leptons are
analyzed separately and are not included when computing the energy of a
cell.  Jets are defined using a simple algorithm.  One first considers the
``hot'' cells (those with transverse energy greater than $5\gev$).  Cells are
combined to form a jet if they lie within a specified ``distance'' or
``radius'',
$R^2={\nabla\eta}^2+{\nabla\phi}^2$, in $\eta$-$\phi$ space from each
other.  Jets have an energy given by the sum of the energy of each cell in
the cluster and a momentum $\vec p_j$ given by the vector sum of the
momentums of each cell.  The invariant mass of a jet is simply
$M_j^2=E_j^2-{\vec p}_j\cdot{\vec p}_j$.  Our single jet and jet pair
definitions are given in Tables 1(a) and 1(b), respectively, and illustrated in
Fig. 1.

\figure{2.5in}{Illustrates the ``bi-polar'' regions used in
analyzing the
profileof a jet pair.  The jet-jet core and center regions are the dark and
medium shaded areas, respectively, while the ``halo'' region is lightly
shaded.}

We have taken the energy resolution to be perfect, which means that the
only resolution effects are caused by the lack of spatial resolution due to
the cell size.  We will examine carefully the effect of energy resolution and
cell size in a forthcoming publication [3].  However, cell size effects tend to
dominate over energy resolution effects, and we have taken rather coarse
cells for this analysis.  The SDC detector at the SSC will have considerably
better spatial resolution with $\De\eta\De\phi=0.05\times3^\circ$ [4].
Thus, we expect that smaller cells will, if anything, improve on what we are
able to do here, even after one includes reasonable energy resolutions.

\vskip 0.2in
\subhead{Lepton Trigger}

The first cut on the data is made by demanding that the event contain {\it at
least} two isolated high transverse momentum leptons ($\ell^\pm=e^\pm$
or $\mu^\pm$) in the central region as follows:

\itemitem{$\bullet$} $P_T(\ell^\pm)>50\gev$, $|\eta(\ell^\pm)|<2.5$.

\noindent Isolated leptons are defined by demanding that the total
transverse energy within a distance $R_\ell$ of the lepton in $\eta$-$\phi$
space be less than $E_T^\ell({\rm max})$.  For this analysis

\itemitem{$\bullet$} $R_\ell=0.2$, $E_T^\ell({\rm max})=5\gev$.

\noindent Lepton pairs ($e^+e^-$ and $\mu^+\mu^-$) are constructed for
the events that survive this first cut.  The pairs are ordered according to
their invariant mass, with pair $\#1$ having the mass closest to the $Z$
boson and pair $\#2$ being the second closest, \etc.  Finally, the event is
rejected unless at least one lepton pair satisfies the following:

\itemitem{$\bullet$}  $P_T(\ell^+\ell^-)>200\gev$.

\figtbl{4.5in}{The 800 GeV Higgs$\to ZZ$ signal and the ``ordinary'' QCD
$Z+$jets background at the SSC energy of $40\tev$.  Results are given (a)
without and (b) with jet-jet profile cuts.  The enhancement factor is defined
to be the percentage of signal ($ZZ\to \ell^+\ell^-jj$ decay mode) divided
by the percentage of background surviving a given set of cuts.}

\noindent Table 2(a) shows that for an $800\gev$ Higgs at the SSC,
roughly $903$ events per year pass this cut for the full decay mode
$ZZ\to$ all.  Here the SSC integrated luminosity for one year is taken to be
$10^4$/pb.  The overall Higgs$\to ZZ$ rate is $11,085$ events per year
with approximately $8\%$ surviving this lepton cut.  About $62\%$ of the
Higgs$\to ZZ\to \ell^+\ell^-q\qbar$ mode pass this cut.  The cross section
for producing a single $Z$ boson at large transverse momentum via the
ordinary QCD subprocesses $qg\to Zq$, $\qbar g\to Z\qbar$, and
$q\qbar\to Zg$ is enormous compared to the Higgs cross section.  As
Table 2(a) shows, $114,662$ events per year of the ordinary QCD
$Z+$jets background survive this lepton cut.

This high transverse lepton pair cut is, of course, crucial.  The transverse
momentum spectrum of the single $Z$ QCD background falls off rapidly,
while for the heavy Higgs the signal is peaked at about half the mass of the
Higgs.  Here one wants to take as large of a cut on $P_T(\ell^+\ell^-)$ as
possible without loosing too much of the signal.  However, even with this
cut, the background is still more than $100$ times the signal!

\vskip 0.2in
\subhead{Jet Pair Selection}
The jet topology of events with at least one large transverse momentum
lepton pair is analyzed by first examining only jet cores (\ie narrow jets of
size $R({\rm core})$).  Here one includes only those jet cores satisfying,

\itemitem{$\bullet$} $E_T({\rm jet\ core})>50\gev$, $|\eta({\rm jet\
core})|<3$.

\noindent In an attempt to find the two jets produced by the hadronic decay
of the large transverse momentum $Z$ boson, jet pairs are formed by
demanding that the distance between the two jet cores in $\eta$-$\phi$
space, $d^2_{jj}=(\eta_1-\eta_2)^2+(\phi_1-\phi_2)^2$, be less than $1.0$.
Namely,

\itemitem{$\bullet$} $d_{jj}$(jet-jet cores)$<1.0$.

\noindent In addition, the jet-jet cores are required to satisfy

\itemitem{$\bullet$} $P_T^{jj}>200\gev$, $|\phi_{jj}-\phi_{ll}|>90^\circ$,

\noindent where $P_T^{jj}$ is the total transverse momentum of the core
jet pair and $\phi_{jj}-\phi_{ll}$ is the azimuthal angle between the leading
lepton pair and the core jet pair.  The jet pair is required to be in the
opposite hemisphere (or ``away-side'') from the lepton pair.  If more than
one jet pair meets all of these requirements than the pair with the largest
total transverse energy is selected.

Table 2(a) shows that of the $903$ Higgs events passing the lepton cut
$52\%$ of the full $ZZ\to$ all mode and $71\%$ of the $ZZ\to \ell^+\ell^-
jj$ channel also pass the jet pair selection criteria.  The full $ZZ\to$ all
mode, of course, contains the $l^+l^-l^+l^-$ and the $l^+l^-\nu\nubar$
channels.  Here we could remove the four charged lepton events, but
instead we analyze simultaneously both the $ZZ\to l^+l^-jj$ and the
``gold-plated'' $ZZ\to l^+l^-l^+l^-$ events.  The four lepton invariant mass
is constructed by taking the leading two lepton pairs.  At this energy and
Higgs mass and with our lepton cuts there are $27$ ``gold-plated'' events
($22$ from the signal and $5$ from the $ZZ$ continuum)..

Unfortunately, $16\%$ of the ordinary QCD $Z+$jets background events
that survive the lepton cuts have an opposite hemisphere jet pair meeting
the selection criteria.  Defining an ``enhancement factor'' as the percentage
of signal ($ZZ\to \ell^+\ell^-jj$ channel) divided by the percentage of
background surviving this cut, one arrives at an enhancement of
$0.71/0.16$ or about $4$.  One might have expected to do better at this
stage.  However, once we required that the $Z$ boson to have a large
transverse momentum, we forced the background to have a large $P_T$
away-side quark or gluon jet.  This away-side parton often fragments via
gluon bremsstrahlung into multiple away-side jets which then survive the
selection criteria.

\vskip 0.2in
\subhead{Invariant Mass Cuts}
The invariant mass, $M_{jj}({\rm full})$, is constructed by using the full
jet pair defined in Table 1(b) and illustrated in Fig. 1.  In particular one
uses
all cells within a ``distance'' $R(halo)$ of either of the jet cores.  The aim
here is, of course, to reconstruct the invariant mass of the $Z$ boson.
However, this full jet-jet invariant mass will {\it only} be used in the event
selection.  The Higgs mass will be reconstructed by setting
$M_{jj}=M_Z$. At this stage, events are rejected unless the full jet-jet
mass satisfies:

\itemitem{$\bullet$} $81<M_{jj}({\rm full})<101$.

\noindent Similarly, the leading lepton pair invariant mass, $M_{\ell\ell}$,
must satisfy:

\itemitem{$\bullet$} $81<M_{\ell\ell}<101$.

\noindent As can be seen from Table 2(a), about $33\%$ of the Higgs
events in the full $ZZ\to$ all mode and $45\%$ of the $ZZ\to \ell^+\ell^-jj$
channel passing {\it both} the lepton cut and the jet pair selection have
$M_{jj}$ within $10\mev$ of the $Z$ boson mass.  For the $ZZ\to
\ell^+\ell^- jj$ Higgs channel, about $63\%$ of the jet pairs have a mass in
this range.  On the other hand, only about $15\%$ of the background jet
pairs have a mass in this range.  About $2.5\%$ of the QCD $Z+$jets
background events surviving {\it both} the lepton cut and the jet pair
selection have a full jet pair invariant mass within $10\mev$ of the $Z$
boson mass.  This corresponds to an enhancement factor of about $18$.

\vskip 0.2in
\subhead{Reconstructing the Higgs Mass}
The Higgs invariant mass is constructed from the momentum vectors of the
two charged leptons and the momentum vector of the jet-jet pair as
follows:
$$
M^2=(E_{\ell^+}+E_{\ell^-}+E_{jj})^2-({\vec p}_{\ell^+}+{\vec
p}_{\ell^-}
+{\vec p}_{jj})^2 \ ,$$
where
$$
E_{jj}^2={\vec p}_{jj}\cdot{\vec p}_{jj}+M_z^2\ .$$

\noindent The mass of a jet is not a well defined quantity since it depends
on the soft particles.  The momentum vector of a jet is better defined and is
determined primarily by the core cells.  Thus, in constructing the Higgs
mass we use the momentum vector of the jet-jet pair but {\it not} the jet-jet
pair mass.  The mass of the jet-jet pair is set equal to the mass of the $Z$
boson.

At this stage, there are $245$ Higgs events and $1,162$ QCD background
events per year at the SSC within $150\gev$ of the true Higgs mass of
$800\gev$, corresponding to an enhancement factor of about $37$ (see
Table 2(a)).  There is also an unavoidable contribution from the $ZZ$
continuum of about $24$ events per year ({\it not shown in Table 2}).
With this enhancement, the $Z+$jets background is still roughly $5$ times
the signal.

\figure{3in}{Shows the fraction of transverse energy within the jet-jet halo
region, $F_{E_T}=E_T($jet-jet halo$)/E_T($full jet-jet$)$, for the Higgs
signal ($ \ell^+\ell^-jj$ mode) and for the $Z+$jets background.  Both
signal and background have passed the lepton cuts, the jet pair selection,
and have $81<M_{jj}({\rm full})<101\gev$.}

\vfill
\head{III. Jet-Jet Profile Analysis}

Our profile analysis of the jet-jet system is accomplished by examining the
``bi-polar'' regions shown in Fig.~1.  The jet-jet system is divided into three
regions.  The first region is the jet-jet core, corresponding to cells whose
centers lie within a ``distance'' $R({\rm core})$ in $\eta$-$\phi$ space of
{\it either} jet.  The jet-jet center region corresponding to cells whose
centers lie within $R({\rm center})$ of {\it either} jet and the full jet-jet
pair region is all the cells whose centers lie within $R({\rm halo})$ of {\it
either} jet.  Cells are {\it not} double counted.  For example, a cell may lie
in the center region of both jets, nevertheless it is counted just once.  The
jet-jet halo region corresponds to cells whose centers lie between $R({\rm
center})$ and $R({\rm halo})$ of {\it either} jet.  These regions are used
to define observables that can differentiate between the jet pairs that
originate from the hadronic decay of a $Z$ boson in the decay of the Higgs
signal and the jet pairs that result from gluon bremsstrahlung from the
recoil parton in large $P_T$ single $Z$ production (\ie the background).

In this analysis we use the following two observables to distinguish signal
from background.  The first is the fraction of the full jet pair transverse
energy that occurs in the jet-jet halo region:
$$
F_{E_T}=E_T(\hbox{jet--jet\ halo})/E_T(\hbox{full\ jet--jet}).
$$
The second observable measured the invariant mass shift from the jet-jet
cores to the full jet pair:
$$
\Delta M=M(\hbox{full\ jet--jet})-M(\hbox{jet--jet\ cores}).
$$
These two observables measure how transverse energy and mass,
respectively, are deposited around the jet-jet cores.

Fig.  2 shows the halo $E_T$ fraction for the Higgs signal and the
$Z+$jets background for events that have survived the lepton cuts, the jet
pair selection, and have $81<M_{jj}({\rm full})<101\gev$.  The
background clearly has more debris in the halo region than the signal.  It is
not surprising to find more $E_T$ surrounding the jet-jet cores in the
$Z+$jets background than in the Higgs signal.  For the signal, the jet pair
arise from the $q\qbar$ decay of a large transverse momentum $Z$ boson.
The $Z$ boson is a color singlet and does not radiate gluons during flight.
On the other hand, the large $P_T$ away-side recoil quarks or gluons in
the single $Z$ background are not color singlets and produce addition
gluons via bremsstrahlung.  These radiated gluons deposit transverse
energy
around the jet-jet cores.  Furthermore, the two jets in the signal originate
from a $q\qbar$ pair, whereas the background is usually a quark-gluon pair
or an antiquark-gluon pair.  Because of the large amount of glue in the
incident protons at the $x_T$ values probed by the SSC, the dominate
subprocesses for large transverse momentum single $Z$ bosons are $qg\to
Zq$ ($55\%$) and $\qbar g\to Z\qbar$ ($37\%$).  The away-side recoil
quark or antiquark radiates a gluon yielding a quark or antiquark plus a
gluon.  However, a gluon produces more gluon radiation than a quark or
antiquark resulting in a different jet-jet profile for the background $qg$ or
$\qbar g$ system relative to the signal $q\qbar$ system.

\figure{3in}{Shows the mass shift, $\Delta M=M(\hbox{full jet--jet})-
M(\hbox{jet--jet cores})$ , for the Higgs signal ($ \ell^+\ell^-jj$ mode) and
for the $Z+$jets background.  Both  signal and background have passed
the lepton cuts, the jet pair selection, and have $81<M_{jj}({\rm
full})<101\gev$.}

Fig. 3 shows the mass shift $\Delta M$ for the Higgs signal and the
$Z+$jets background for events that have survived the lepton cuts, the jet
pair selection, and have $81<M_{jj}({\rm full})<101\gev$.  On the
average, the mass shift is larger for the background.  The background has
more mass located around the jet-jet cores for the same reason it has more
transverse energy in the halo.  Because of this, the two observables are not
completely ``orthogonal''.  Nevertheless, both the halo $E_T$ and the mass
shift are important observables for differentiating signal from
background\footnote{${}^\dagger$}{By studying the effect of both profile
cuts on the {\it same} region about the jet-jet centers, we have verified that
these cuts are significantly different [3].}.  They can be used together to
preferentially select the signal over the background

\vfill\eject
\vskip 0.2in
\subhead{Profile Cuts}
One can make cuts on the halo $E_T$ and the mass shift in a variety of
ways. In our big paper [3], we discuss the optimization of these cuts.  Here
we illustrate the power of profile cuts by taking,

\itemitem{$\bullet$} $F_{E_T}<1.5\%$ and $\Delta M<15\gev$ .

\noindent The result of making these cuts is shown in Table 2(b).  Now,
there are approximately $106$ Higgs events and $56$ QCD background
events per year at the SSC within $150\gev$ of the Higgs mass (and $10$
events from the $ZZ$ continuum). This corresponds to an overall
enhancement factor of around $300$, and is about a factor of $8$ better
than without jet-jet profile cuts!  Of course, the signal has been reduced by
more than $50\%$. On the other hand, one does not need to cut as hard as
we have done here. Furthermore, jet-jet profile cuts can be used in
conjunction with other cuts, such as forward jet tagging [5], and cuts that
make use of the longitudinal polarization of the $Z$-bosons coming from
Higgs decay [6].

\figure{3in}{Shows the reconstructed Higgs mass of an $800\gev$ Higgs
produced in proton-proton collisions at the SSC energy of $40\tev$.  The
plot corresponds to the number of events in a $100\gev$ bin per SSC year
for the {\it sum} of the Higgs signal, the $ZZ$ continuum, and the
$Z$+jets background that survive the lepton cuts and  the jet-jet pair
selection {\it with} jet-jet profile cuts.  For comparison, the
``gold-plated''
$\ell^+\ell^-\ell^+\ell^-$ events passing the lepton cuts are also shown.}

\figure{3in}{Shows the reconstructed Higgs mass of an $800\gev$ Higgs
produced in proton-proton collisions at the SSC energy of $40\tev$.  The
plot corresponds to the number of events in a $100\gev$ bin per SSC year
for the {\it sum} of the Higgs signal, the $ZZ$ continuum, and the
$Z$+jets background that survive the lepton cuts and  the jet-jet pair
selection {\it without} jet-jet profile cuts.}

Fig. 4 shows the reconstructed Higgs mass of an $800\gev$ Higgs
produced in proton-proton collisions at the SSC energy of $40\tev$ after
the jet-jet profile cuts have been employed.  The plot corresponds to the
number of events in a $100\gev$ bin per SSC year for the {\it sum} of the
Higgs signal, the $ZZ$ continuum, and the $Z$+jets background.  The
contribution to the sum from the signal and the background are shown by
the light and dark shaded regions, respectively, with the hatched area
representing the $ZZ$ continuum contribution.  As a result of all the cuts,
the background peaks in the mass region from $650$-$1150$.
Nevertheless, the signal shows up as a peak above the background and
with a rate that is still about $5$ times that of the ``gold-plated'' $l^+l^-
l^+l^-$ mode.   For comparison, Fig. 5 shows the reconstructed Higgs
mass {\it without} jet-jet profile cuts.

\vfill\eject
\head{IV. Summary and Conclusions}

We have devised a method that can help to distinguish the two jet system
originating from $q\qbar$ the decay of a color singlet $Z$ boson from a
random jet pair coming from the ``ordinary'' QCD gluon bremsstrahlung of
colored quarks and gluons.  Two observables are defined that measure how
transverse energy and mass, respectively, are distributed around the jet-jet
system.  The procedure can be summarized by the following series of
selections and cuts:

\itemitem{$\bullet$} Lepton pair trigger.
\itemitem{$\bullet$} Jet pair selection.
\itemitem{$\bullet$} Jet-jet profile cuts.
\itemitem{$\bullet$} Jet-jet invariant mass cuts.

\noindent The invariant mass of the jet-jet pair is used {\it only} in the
selection of events, the Higgs mass is reconstructed from the momentum of
the jet pair with $M_{jj}$ set equal to $M_z$.  We are able to obtain
signal to background enhancements greater than $100$. With
enhancements this large, the Higgs stands out in the invariant mass plot as
a definite peak over the background (see Fig. 4).  The method works
equally well for $600\gev$ Higgs and we are currently studying Higgs
masses of $200$ and $400\gev$ [3].

Jet-jet ``shape cuts'' have been considered previously.  The two which are
most similar to our method are the ``elongation'' of the jet pair, and the
chi--squared shape [7].  However, jet-jet profile cuts are better able to
differentiate the between the underlying color structure of the Higgs-boson
events and the QCD background, because they measure the activity in the
``halo'' regions of the jet pair, rather than just examining the overall pair
shape.

Hadronic multiplicity cuts [8] rely upon the same difference of color
structure between signal and background.  However, observation of
hadronic multiplicity requires a high performance silicon $\mu$strip tracker
[9]. The final design status of this component of the SSC detectors, and its
final performance at the SSC, is not clear [1,2].  Furthermore, hadronic
multiplicity is not estimated reliably by the Monte-Carlo event generators
and depends on the jet fragmentation model employed [10].

We believe that further improvements in enhancing the Higgs$\to ZZ\to
\ell\ell jj$ signal over the $Z$+jets background can be made by combining
jet-jet profile cuts with other ``orthogonal'' cuts like forward jet tagging
[5], and cuts that make use of the longitudinal polarization of the
$Z$-bosons from Higgs decay [6].  Also, we think we can improve on the
jet pair selection criteria [3].  Furthermore, our method also works for
$W$ bosons and should help clean up the Higgs$\to WW\to \ell\nu jj$
signal as well.

\vskip0.2in
\centerline{\bf Acknowledgments}
We thank Frank Paige for his help with ISAJET and for his useful
comments during his visit to Florida.

\vskip 0.2in
\noindent {\bf References}
\singlespace
\item{1.} SDC Letter of Intent, SSC Preprint SSCL-SR-1153A (1990),
 and references therein.
\item{2.} GEM Letter of Intent, SSC Preprint SSCL-SR-1184 (1991),
Hong Ma, GEM internal report (1991).
\item{3.} {\it Enhancing the Higgs Signal at the SSC}, R.~D.~Field and
P.~A.~Griffin, in preparation.
\item{4.} SDC Technical Design Report, SSC Preprint Draft-SDC-92-201
(1992).
\item{5.} R.N.~Cahn, S.D.~Ellis, R.~Kleiss, W.J.~Stirling, \prd,
35,1626,1987.\hfil\break
 R.~Kleiss and W.J.~Stirling, \pl 200B,193,1988.
\item{6.} J.F.~Gunion and M.~Soldate, \prd,34,826,1986.
\item{7.} G.~Alverson, {\it et. al.}, p.~114, in {\it Proceedings of the
1986 Summer Study on Physics at the SSC}, Snowmass, Colorado,
1986, edited by Rene Donaldson and Jay Marx, (Division of Particles and
Fields of the American Physical Society, 1986).
\item{8.} J.F.~Gunion, {\it et. al.}, \prd, 40, 2223, 1989.
\item{9.} H.F.-W. Sadrozinski, A.Seiden, A.J.~Weinstein, {\sl Nucl.
Instrum. Methods} {\bf A277}, 92 (1989).
\item{10.}H.F.-W. Sadrozinski, A.Seiden, A.J.~Weinstein, in {\it
Proceedings of
the 1988 Summer Study on High Energy Physics in the 1990's},
Snowmass,
Colorado,
1988, edited by Sharon Jensen, (World Scientific, Singapore, 1989).

\endit
\end